\documentclass[prb,superscriptaddress,twocolumn]{revtex4-2}
\usepackage{graphicx}
\usepackage{times,amsmath}
\usepackage{epsfig}
\usepackage{color}
\usepackage{graphicx}
\usepackage{dcolumn}
\usepackage{bm}
\usepackage{bookmark}
\usepackage{tabularx}
\usepackage{hyperref}
\usepackage{multirow}
\usepackage{array,mathtools,amssymb,booktabs,makecell}
\hypersetup{colorlinks=true, citecolor=blue, filecolor=blue, linkcolor=blue, urlcolor=blue}

\usepackage{float}
\usepackage{adjustbox}
\usepackage{upgreek}
\usepackage[utf8x]{inputenc}

\makeatletter
\let\saved@includegraphics\includegraphics
\AtBeginDocument{\let\includegraphics\saved@includegraphics}
\renewenvironment*{figure}{\@float{figure}}{\end@float}
\makeatother

\begin{document}

\title{Dual Nature of Magnetism Driven by Momentum Dependent $f$-$d$ Kondo Hybridization}

\author{Byungkyun Kang}
\email[]{bkang@udel.edu}
\affiliation{College of Arts and Sciences, University of Delaware, Newark, DE 19716, USA}

\author{Yongbin Lee}
\affiliation{Ames Laboratory, U.S.~Department of Energy, Ames, IA, 50011, USA}

\author{Liqin Ke}
\affiliation{Ames Laboratory, U.S.~Department of Energy, Ames, IA, 50011, USA}

\author{Hyunsoo Kim}
\affiliation{Department of Physics, Missouri University of Science and Technology, Rolla, 65409, MO, USA}

\author{Myoung-Hwan Kim}
\affiliation{Department of Physics and Astronomy, Texas Tech University, Lubbock, Texas 79409, USA}

\begin{abstract}
Intricate nature of magnetism in uranium-based Kondo lattices is a consequence of correlations between U-5$f$ and conduction electrons. 
Using linearized quasiparticle self-consistent GW plus dynamical mean-field theory, we demonstrate a crossover from incoherent to coherent $f$-$d$ Kondo cloud in the paramagnetic phase of UTe$_2$ with reduced volumes, USbTe and USbSe. As the transition occurs, we observe an augmented $f$-$d$ coherence and Pauli-like magnetic susceptibility, with a substantial frozen magnetic moment of U-5$f$ persisting. 
We show that momentum dependent $f$-$d$ hybridization is responsible for the magnetic moments arising from the renormalized $f$ electrons' van Hove singularity.
Our findings provide a unique perspective to explain the dual nature of magnetism and the long-range magnetic ordering induced by pressure in UTe$_2$.


\end{abstract}

\maketitle

Bewildering behavior of U-5$f$ electrons in uranium-based compounds, extending from localisation to mobility and dualism~\cite{schoenes_prb1996,jooseop_prl2018,andrea_pnas2020}, is further perplexed by Kondo screening by conduction electrons, leading to the advent of unprecedented phenomena in a variety of Kondo lattices. Amorese et al.\cite{andrea_pnas2020} elucidated the dual nature of local atomic-like multiplet states in Pauli-paramagnetic UFe$_2$Si$_2$, which has been known to exhibit Kondo-like behaviour at low temperatures~\cite{szytula_jmmm1988}. USbTe, a Kondo lattice ferromagnet, has been found to possess a large anomalous Hall conductivity, which is suggested to be due to the intrinsic Berry curvature hosted by the Kondo hybridization between the local magnetic moment of U-5$f$ and the conduction electrons~\cite{byung_usbte}. UTe$_2$, a heavy fermion superconductor, has been reported to lack long-range magnetic ordering~\cite{Aoki2021review}, in contrast to other uranium compounds, such as USbTe and USbSe ($T_\textrm{C}$$\sim$127 K)~\cite{kaczorowski_jmmm1995}. Nonetheless, UTe$_2$ has been found to exhibit pressure-induced long-range magnetic ordering~\cite{thomas_scadv2021,dexin_jpsj2021,braithwaite_comphys2019,sheng_prb2020}. Li et al.~\cite{dexin_jpsj2021} demonstrated that increasing pressure leads to a diminution of magnetic moments and long-range magnetic ordering. The manifestly disparate behaviours exhibited in U-5$f$ systems reflect a profound bond between U-5$f$ and its surroundings, and are not easily explicable by existing theories.

Theoretical studies have proposed the presence of orbital-selective Mott phases (OSMP) to explain the duality observed in multi-band correlated systems~\cite{anisimov_epjb2002, evgeny_prl2022,fabian_prl2022}. This phenomenon is characterized by the coexistence of localized electrons in certain orbitals and itinerant electrons in other orbitals.
The OSMP has been employed to explain the dual nature of Fe-based compounds. 
For example, Kim et al.~\cite{minsung_prb2020} reported that the Mott transition in FePS$_3$ can be orbital selective, with the $t_{2g}$ states undergoing a correlation-induced insulator-to-metal transition while the $e_{g}$ states remain gaped under pressure.
However, no evidence of the occurrence of the OSMP has been found in the 5$f$ system.

Unlike $4f$ electrons in the lanthanide elements, which are typically localized within the Mott physics~\cite{kang_ndnio2}, the degree of localization of $5f$ electrons in the actinide elements is strongly dependent on the crystal structure~\cite{schoenes_prb1996}, crystal electric field, and spin-orbit coupling (SOC).
These factors can affect the hybridization channels, leading to a varying screening of the U-5$f$ magnetic moments~\cite{leonid_prm2023}, thereby altering the magnetization mechanism depending on whether Kondo hybridization is coherent or incoherent.
According to the mutichannel Kondo model~\cite{multichannel,nozieres1980kondo}, 
Kondo systems can be classified into three types (under, fully, over-screened) based on the local magnetic moment raised by impurity spin $S$ and the number of conduction electron channels, $n$. In the under-screened case ($n < 2S$), $S$ is partially screened at low temperatures, potentially allowing for long-range magnetic ordering due to the Ruderman Kittel-Kasuya-Yosida (RKKY) interaction.
An example of this is the Anderson lattice model of uranium monochalcogenides, where UTe is modelled with $S = 1$ and $n < 2$~\cite{perkins_prb2007,thomas_prb2011}.
However, the RKKY mechanism may not be applicable to coherent U-5$f$ Kondo lattices, as the Kondo coupling and the RKKY interaction favor different ground states~\cite{hilbert_rmp2007,doniach_phsica1977}.

In a Bloch electrons system, van Hove singularity (VHS)~\cite{van1953} can be attributed as a source of magnetisation. 
As the Fermi energy approaches a VHS of the electronic density of states (DOS), the DOS diverges, allowing for weak interactions to have a significant influence on the electronic behavior. This can result in instabilities in charge and spin susceptibilities, leading to substantial enhancement of ferromagnetism~\cite{fleck_prb1997,hlubina_prl1997} and antiferromagnetism~\cite{lin_prb1987}.
The VHS has been associated with the local magnetic moment in a transition metal~\cite{hausoel_ncomm2017} and graphene bilayer~\cite{liu_prb2019}.
By using density functional theory (DFT) combined with dynamical mean field theory (DMFT), Hausoel et al.~\cite{hausoel_ncomm2017} recently investigated the paramagnetic phase spectrum of Ni metal. Their results revealed a van Hove magnet at the L point of the Brillouin zone, characterized by a large effective mass and temperature-dependent magnetic susceptibility.
This feature has not been previously observed in U-5$f$ Kondo lattice.

In this work, we scrutinized the alteration from disordered to ordered $f$-$d$ Kondo hybridization in the paramagnetic phase of Kondo lattices of UTe$_2$, UTe$_2$ with reduced volume, USbTe and USbSe. 
We discovered that UTe$_2$ with reduced volume and USbSe are in the coherent regime, exhibiting Pauli-like magnetic susceptibility and dispersive bands arising from the hybridization of U-6$d$ and renormalized U-5$f$ electrons. This leads to delocalized U-5$f$ electrons, i.e. Bloch-like quasiparticle states, and a remarkably significant U-5$f$ magnetic moment. We ascribed this duality of Pauli-like magnetic susceptibility and local magnetic moments to momentum reliant coherent $f$-$d$ hybridization, with the VHS being the origin of the magnetic moment. 
This feature was employed to investigate the emergence of long-range magnetic ordering in UTe$_2$ under pressure.

\textit{Coulomb interaction tensor.} 
\begin{figure}[ht]
\centering
\includegraphics[width=0.5 \textwidth]{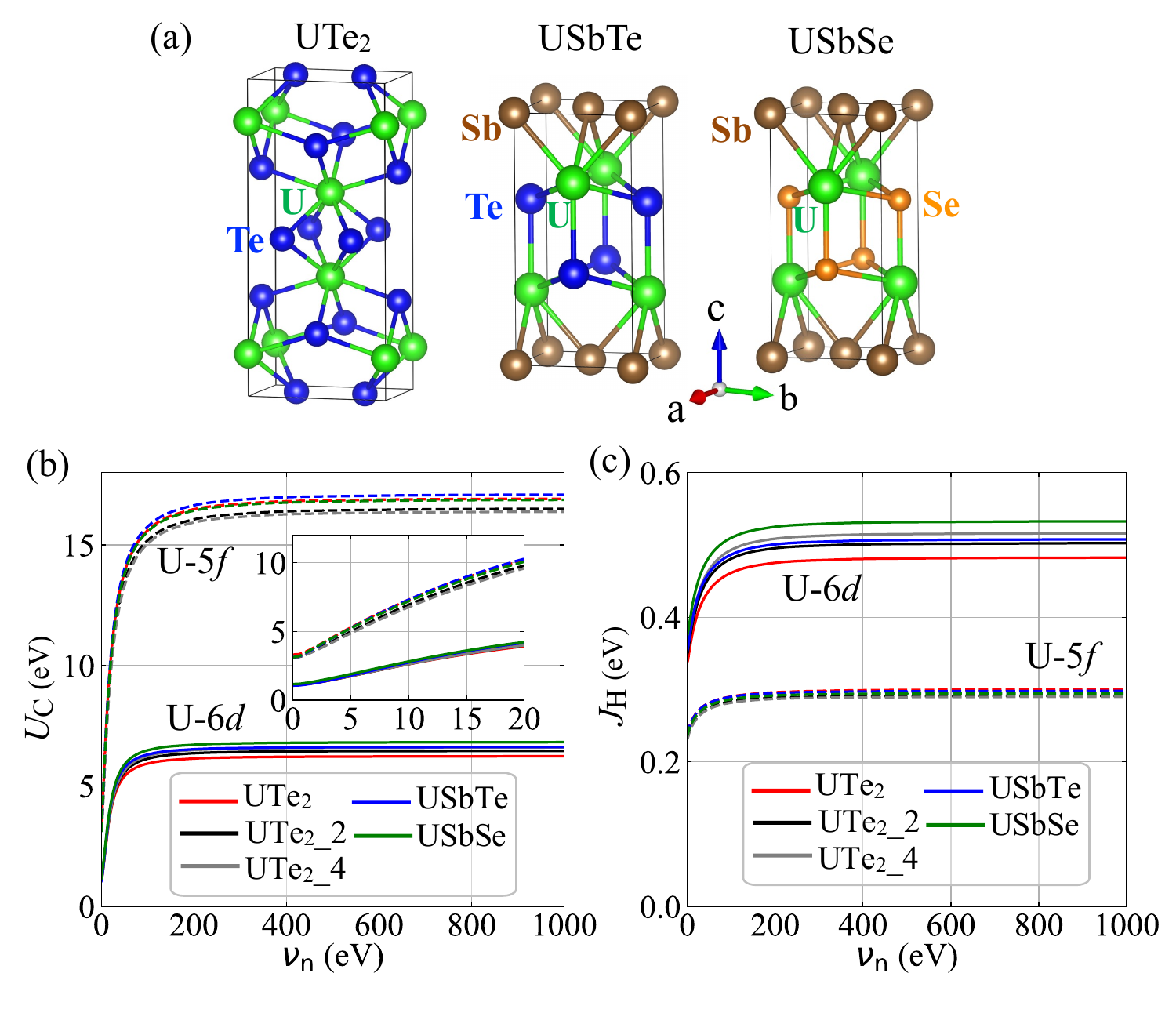}
\caption{\label{Fig_u}\
(a) Crystal structure of UTe$_2$, USbTe, and USbSe.
Calculated onsite (b) Coulomb interaction $U_{\textrm{C}}$ and (c) exchange interaction $J_{\textrm{H}}$ for U-5$f$ and U-6$d$ with inclusion of SOC as a function of $\nu_n = 2n\pi/\beta$ bosonic frequencies. In (b), inset shows magnified views of $U_{\textrm{C}}$ in low frequency range.}
\end{figure}
The crystal structures of the orthorhombic phase (Immm) UTe$_{2}$~\cite{ikeda2006single} and the tetragonal phase (P4/nmm) USbTe and USbSe~\cite{hulliger_jlcm1968} are depicted in Fig.~\ref{Fig_u} (a). 
To investigate the impact of pressure on the electronic structure of UTe$_2$, structures with a decreased volume of UTe$_2$ were generated by reducing the experimental lattice parameters~\cite{ikeda2006single} by 2 $\%$ (UTe$_2$$\_2$) and 4 $\%$ (UTe$_2$$\_4$). 
The constrained random phase approximation~\cite{aryasetiawan2004frequency} was employed to calculate the onsite Coulomb interaction $U_\textrm{C}$ and exchange interaction $J_{\textrm{H}}$, for U-5$f$ and U-6$d$ orbitals, taking into account SOC.
Fig.~\ref{Fig_u} (b) and (c) show that both $U_\textrm{C}$ and $J_\textrm{H}$ increase and reach their unscreened values at high frequencies. The static $U_{\textrm{C}}$ of U-6$d$ and U-5$f$ orbitals were found to be comparable across the compounds studied. 
However, $U_{\textrm{C}}$ of U-5$f$ is larger than that of U-6$d$, while $J_{\textrm{H}}$ for U-5$f$ is smaller than that of U-6$d$.


\begin{table*}[]
\caption{Calculated  electron occupation of U-5$f$ and U-6$d$ orbitals in UTe$_{2}$, USbTe, and USbSe at $T=300$ K. U-5$f$ and U-6$d$ orbitals are labelled for convenience in this work. }\label{table_occ}
\scriptsize
\begin{center}
\scalebox{1}{
\begin{ruledtabular}

\begin{tabular}{c|cccccc|cccccccc|cccc|cccccc}

 \multicolumn{1}{c|}{} &
 \multicolumn{14}{c|}{U-5$f$} &
 \multicolumn{10}{c}{U-6$d$} \\
 \hline
  $j$ & \multicolumn{6}{c|}{5/2} & \multicolumn{8}{c|}{7/2} & \multicolumn{4}{c|}{3/2} & \multicolumn{6}{c}{5/2} \\
  \hline
  $j_{z}$ &-2.5& -1.5 & -0.5 & 0.5 & 1.5 & 2.5 & -3.5 & -2.5 & -1.5 & -0.5 & 0.5 & 1.5 & 2.5 & 3.5 & -1.5 & -0.5 & 0.5 & 1.5 & -2.5 & -1.5 & -0.5 & 0.5 & 1.5 & 2.5\\
  \hline
  label &$f_{1}$& $f_{2}$ & $f_{3}$ & $f_{4}$& $f_{5}$ & $f_{6}$ & $f_{7}$ & $f_{8}$ & $f_{9}$ & $f_{10}$ & $f_{11}$ & $f_{12}$ & $f_{13}$ & $f_{14}$ &$d_{1}$& $d_{2}$ & $d_{3}$ & $d_{4}$& $d_{5}$ & $d_{6}$ & $d_{7}$ & $d_{8}$ & $d_{9}$ & $d_{10}$\\
  \hline
  UTe$_{2}$ &0.37& 0.31 & 0.36 & 0.36 & 0.30 & 0.36 & 0.03 & 0.03 & 0.03 & 0.03 & 0.03 & 0.03 & 0.03 & 0.03 & 0.21 & 0.23 & 0.23 & 0.21 & 0.17 & 0.19 & 0.19 & 0.19 & 0.19 & 0.17 \\ 
  \hline
    UTe$_{2}$$\_2$ &0.35& 0.30 & 0.36 & 0.35 & 0.30 & 0.35 & 0.03 & 0.03 & 0.03 & 0.03 & 0.03 & 0.03 & 0.04 & 0.03 & 0.21 & 0.22 & 0.22 & 0.21 & 0.17 & 0.17 & 0.19 & 0.19 & 0.17 & 0.17 \\ 
  \hline
    UTe$_{2}$$\_4$ &0.35& 0.27 & 0.33 & 0.34 & 0.27 & 0.34 & 0.04 & 0.04 & 0.05 & 0.05 & 0.05 & 0.05 & 0.04 & 0.04 & 0.21 & 0.23 & 0.23 & 0.21 & 0.18 & 0.18 & 0.20 & 0.20 & 0.17 & 0.18 \\   
  \hline
  USbTe &0.34& 0.34 & 0.33 & 0.33 & 0.35 & 0.35 & 0.03 & 0.03 & 0.03 & 0.03 & 0.03 & 0.03 & 0.03 & 0.03 & 0.23 & 0.22 & 0.22 & 0.23 & 0.18 & 0.17 & 0.18 & 0.18 & 0.17 & 0.18 \\ 
  \hline
  USbSe &0.34& 0.35 & 0.33 & 0.33 & 0.36 & 0.35 & 0.04 & 0.03 & 0.03 & 0.04 & 0.04 & 0.03 & 0.04 & 0.04 & 0.20 & 0.20 & 0.20 & 0.20 & 0.15 & 0.16 & 0.17 & 0.17 & 0.16 & 0.15 \\   
\end{tabular}
\end{ruledtabular}}
\end{center}
\end{table*}

\begin{figure*}[ht]
\centering
\includegraphics[width=1.0 \textwidth]{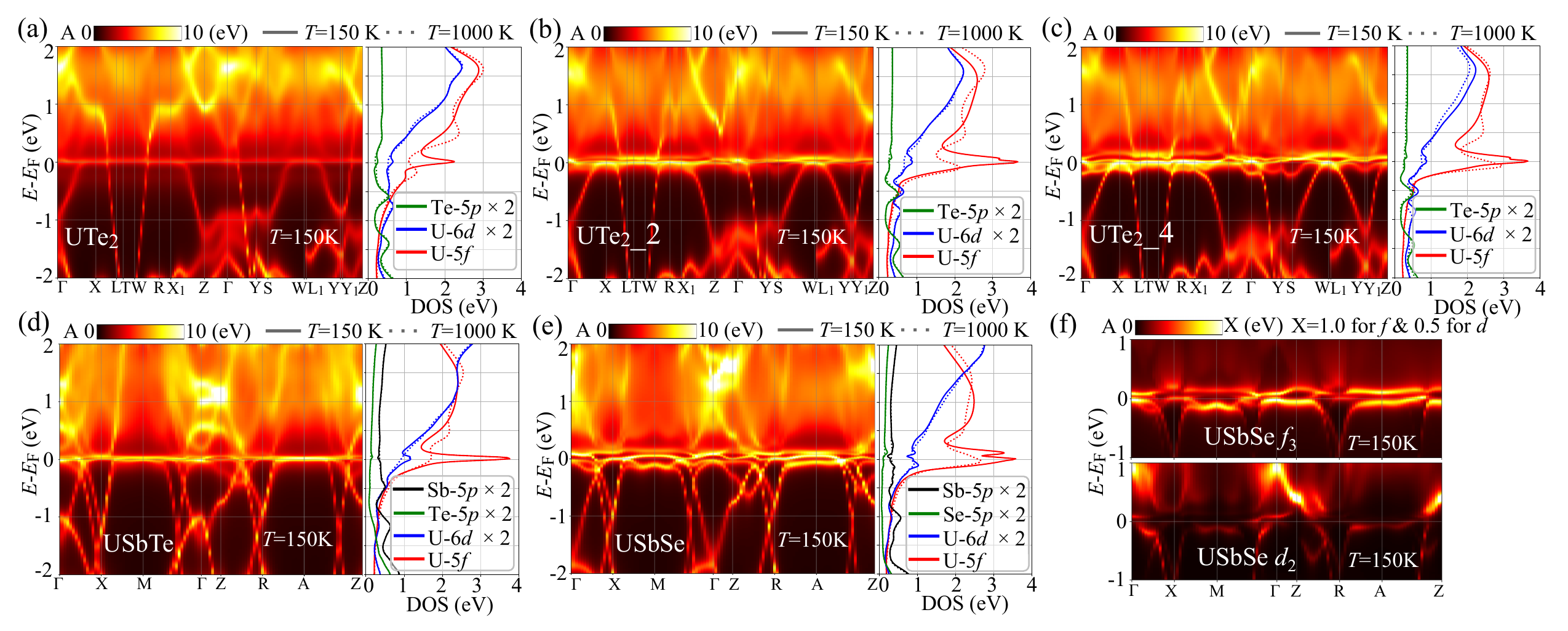}
\caption{\label{Fig_dos} 
Calculated spectral functions and DOS for (a) UTe$_2$, (b) UTe$_2$$\_$2, (c) UTe$_2$$\_$4, (d) USbTe, (e) USbSe. (f) Orbital projected spectral functions for USbSe. $f_3$ ($d_2$) projected spectral function is presented in upper (lower) panel (see the orbital labeling in Table~\ref{table_occ}). In the top of each figure, A denotes spectral weight.
}
\end{figure*}

\textit{Electronic structures.}
The local self-energies of U-5$f$ and U-6$d$ were determined by solving two distinct single impurity models using the continuous time quantum Monte Carlo method within the framework of DMFT~\cite{choi2019comdmft}. The dynamic Coulomb interaction tensors of U-5$f$ and U-6$d$ were employed in this process.
By eliminating the off-diagonal elements in the hybridization functions, the local self-energy is assumed to be diagonal in the spherical harmonics basis. Supplementary Figure 1 (Fig. S1) demonstrates that the off-diagonal elements are relatively small and insignificant.
Further details of the methods used can be found in the supplementary materials.
The electron occupancies of U-5$f$ and U-6$d$ orbitals at 300 K are presented in Table~\ref{table_occ}.
The U-6$d$ orbitals were found to have a spread occupation across all $d$ orbitals, while the U-5$f$ orbitals were split into $j=$ 5/2 and $j=$ 7/2 multiplets, with a significant occupation of the $j=$ 5/2 multiplet.
The total occupations of U-5$f$ orbitals were 2.27, 2.25, 2.26, 2.24, and 2.31 for UTe$_2$, UTe$_2$$\_2$, UTe$_2$$\_4$, USbTe, and USbSe, respectively. 
The U-5$f$ occupation for UTe$_2$ was found to be in accordance with the measured 5$f^{2}$ configuration~\cite{stowe1997uncommon}, indicating a substantial local magnetic moment of U-5$f$ and the presence of an orbital selective Kondo effect~\cite{byung_ute2}.

Figure~\ref{Fig_dos} presents the calculated spectral functions and atomic-orbital projected DOS. 
It is observed that all compounds share a common property of a strong U-5$f$ peak in the vicinity of the Fermi level at low temperature. The temperature dependence of the U-5$f$ peak between 150 K and 1000 K is consistent with that of the U-6$d$ peak, indicating hybridization between U-5$f$ and U-6$d$. 
In contrast, the DOS of other anions of Te, Sb, and Se are small and their temperature dependence are subtle. Furthermore, the presence of flat $f$ bands and kink-like band structures implies that $f$-$d$ Kondo hybridization governs the electronic structure in the vicinity of the Fermi level~\cite{byung_ute2,choi2013observation,kang_ndnio2}. This is in agreement with the experimental observation of Kondo lattice in UTe$_{2}$~\cite{Jiao2020}, USbTe~\cite{usbte_r,byung_usbte}, and USbSe~\cite{usbte_r}.
The calculated Kondo hybridization of USbTe with respect to temperature was found to be in agreement with angle-resolved photoemission spectroscopy (ARPES) measurements~\cite{byung_usbte}. Similarly, the dispersive Kondo resonance peaks near the Fermi level of UTe$_2$ were also observed in the ARPES~\cite{Fujimori2019}.

\textit{Crossover from incoherent to coherent.}
\begin{figure}[ht]
\centering
\includegraphics[width=0.5 \textwidth]{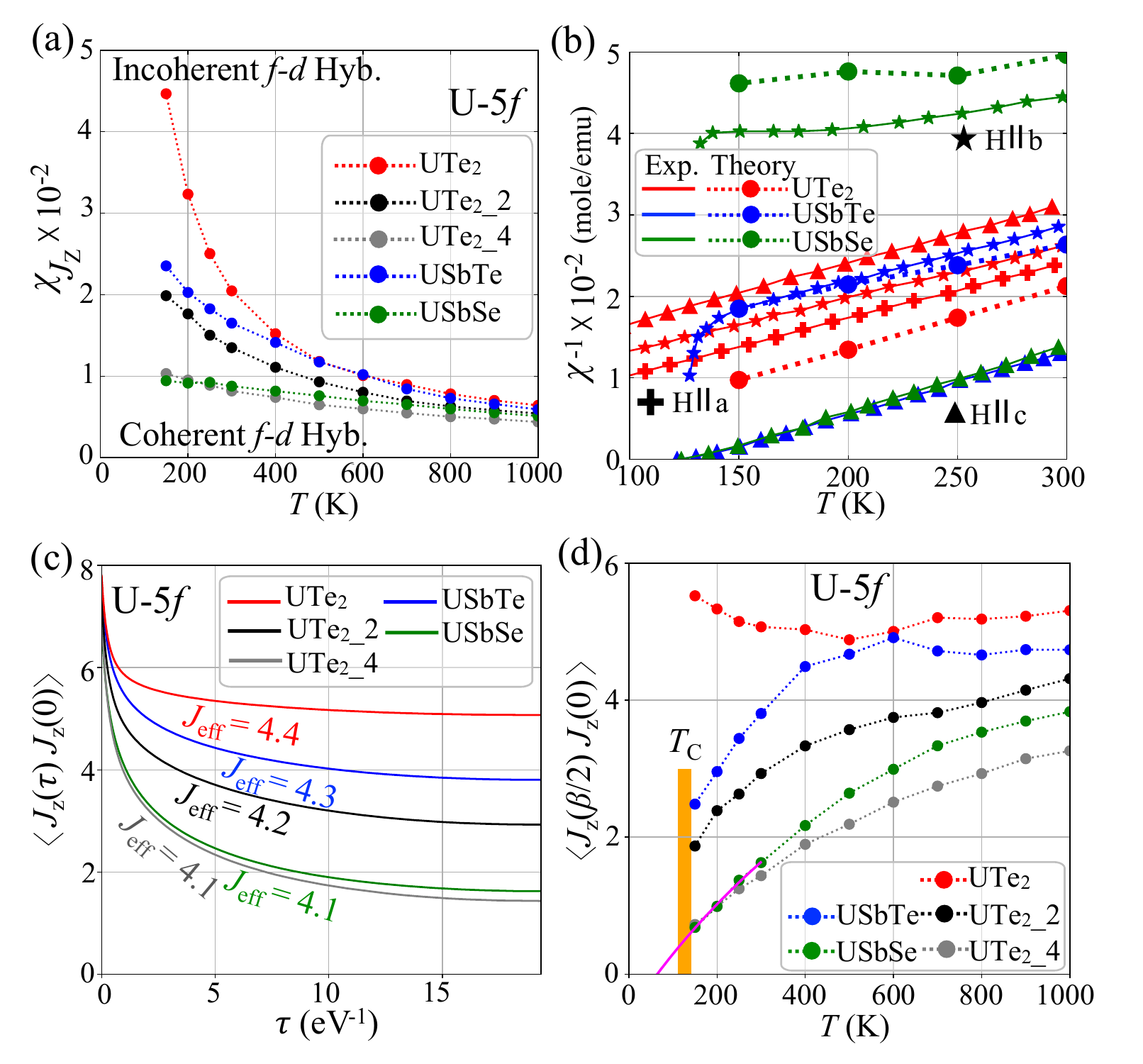}
\caption{\label{Fig_sus} 
(a) Calculated local total angular momentum susceptibility. (b) Comparison of inverse of magnetic susceptibility between calculations and experiments~\cite{kaczorowski_jmmm1995,knafo_comphy2021}. (c) Local angular moment correlation functions in the imaginary time $\tau$. 
Calculated instantaneous $J$ are presented with corresponding colors. (d) Temperature dependent frozen magnetic moments. The bar shows the ferromagnetic transition temperature $T_\textrm{C}=$127 K of USbTe and USbSe~\cite{kaczorowski_jmmm1995}. }
\end{figure}
The local total angular momentum susceptibility, $\chi_{J_Z}$, was calculated as $\int_0^\beta d\tau\langle J_z(\tau)J_z(0)\rangle$. 
Fig.~\ref{Fig_sus} (a) highlights the $\chi_{J_Z}$ of U-5$f$ for UTe$_{2}$, which exhibits a Curie-like behavior. Upon decreasing the volume, UTe$_{2}$$\_4$ exhibits Pauli-like behavior of weak temperature dependence, while USbTe and UTe$_{2}$$\_2$ display an intermediate susceptibility. USbSe, on the other hand, shows Pauli-like behavior. 
At temperatures between 150 and 250 K, a flat $\chi_{J_Z}$ was observed for USbSe which clearly shows in Fig.~\ref{Fig_sus}(b) both experiment and theory.
These results suggest that the magnetic moment on UTe$_{2}$ is strong and resilient, while in USbSe and UTe$_{2}$$\_4$, the electrons from the $s$, $p$, $d$ and $f$ orbitals of neighboring atoms are arranged in an antiparallel manner to the moment of U-5$f$, resulting in the cancellation of net moments~\cite{marianetti_prl2008}.

Comparison of $\chi_{J_Z}$ with experiments~\cite{kaczorowski_jmmm1995,knafo_comphy2021} yielded good agreement, as demonstrated in Fig.~\ref{Fig_sus}(b).
The inverse of the calculated $\chi$=$g\chi_{J_Z}k_{\textrm{B}}/3$ is presented, where $g$ is set to 0.8~\cite{freeman_prb1976,jooseop_prl2018}.
The calculated $\chi^{-1}$ of UTe$_2$, USbTe, and USbSe is found to be in the order of UTe$_2$ $<$ USbTe $<$ USbSe, which is consistent with the experimental $\chi^{-1}$ when a magnetic field is applied along the b axis (H$\parallel$b).
The slope of linear $\chi^{-1}$ of UTe$_2$ is confirmed by both experimental and theoretical results.
For USbTe, both experiment (H$\parallel$b) and theoretical calculations show a linear $\chi^{-1}$ down to 150 K, albeit with a slightly different slope. 
Experiment $\chi^{-1}$ (H$\parallel$b) of USbSe exhibits Pauli-type behavior from approximately 140 to 190 K, followed by Curie-like behavior at higher temperatures. Our calculations further suggest Pauli-type behavior from 150 to 250 K, albeit with a slightly different temperature scale. This implies that USbSe is in the coherent regime with Pauli-type behavior at low temperatures, corroborated by both theory and experiment.

The local angular moment correlation functions $\chi_{J_Z}(\tau)=\langle J_z(\tau)J_z(0)\rangle$ are presented in Fig.~\ref{Fig_sus} (c). The correlation function can be used to assess the degree of magnetic moment localization~\cite{belozerow_prb2023}.
The instantaneous $J$ values of U$^{4+}$ range from 4.4 to 4.1, which is consistent with the $^{3}H_{4}$ Russell Saunders ground state configuration~\cite{freeman_prb1976}.  
Fig. S2 shows that the charge susceptibility $\chi_\textrm{N}=\int_0^\beta d\tau \langle N(\tau) N(0)\rangle$ of U-5$f$ electrons is enhanced in the coherent regime, while the $\chi_{J_Z}(\tau)$ is reduced, as illustrated in Fig.~\ref{Fig_sus} (c).
These indicate that the itinerant character of the U-5$f$ electrons in USbSe and UTe$_{2}\_$4 is significantly enhanced, as further evidenced by the diverging hybridization functions in these materials, as shown in Fig. S3. 
Remarkably, USbSe and UTe$_{2}\_$4 exhibit a saturated value of $\sim$ 2.0 for $\chi_{J_Z}(\tau)$ at $\tau\xrightarrow{} \beta/2$, indicating that the U-5$f$ electrons are localized (not completely screened) and thus a frozen magnetic moment is present. 
Figure~\ref{Fig_sus} (d) shows that the frozen magnetic moment of USbSe does not vanish entirely before the ferromagnetic transition upon cooling.
Our research reveal that UTe$_{2}$ under pressure and USbSe are in the coherent regime.
Notably, USbSe exhibits Pauli-type behavior and a local magnetic moment at low temperature, indicating a dual character of the U-5$f$ electrons, with both localization and itinerancy present.


\textit{Duality driven by momentum dependent $f$-$d$ hybridization.}
\begin{figure*}[ht]
\centering
\includegraphics[width=0.88 \textwidth]{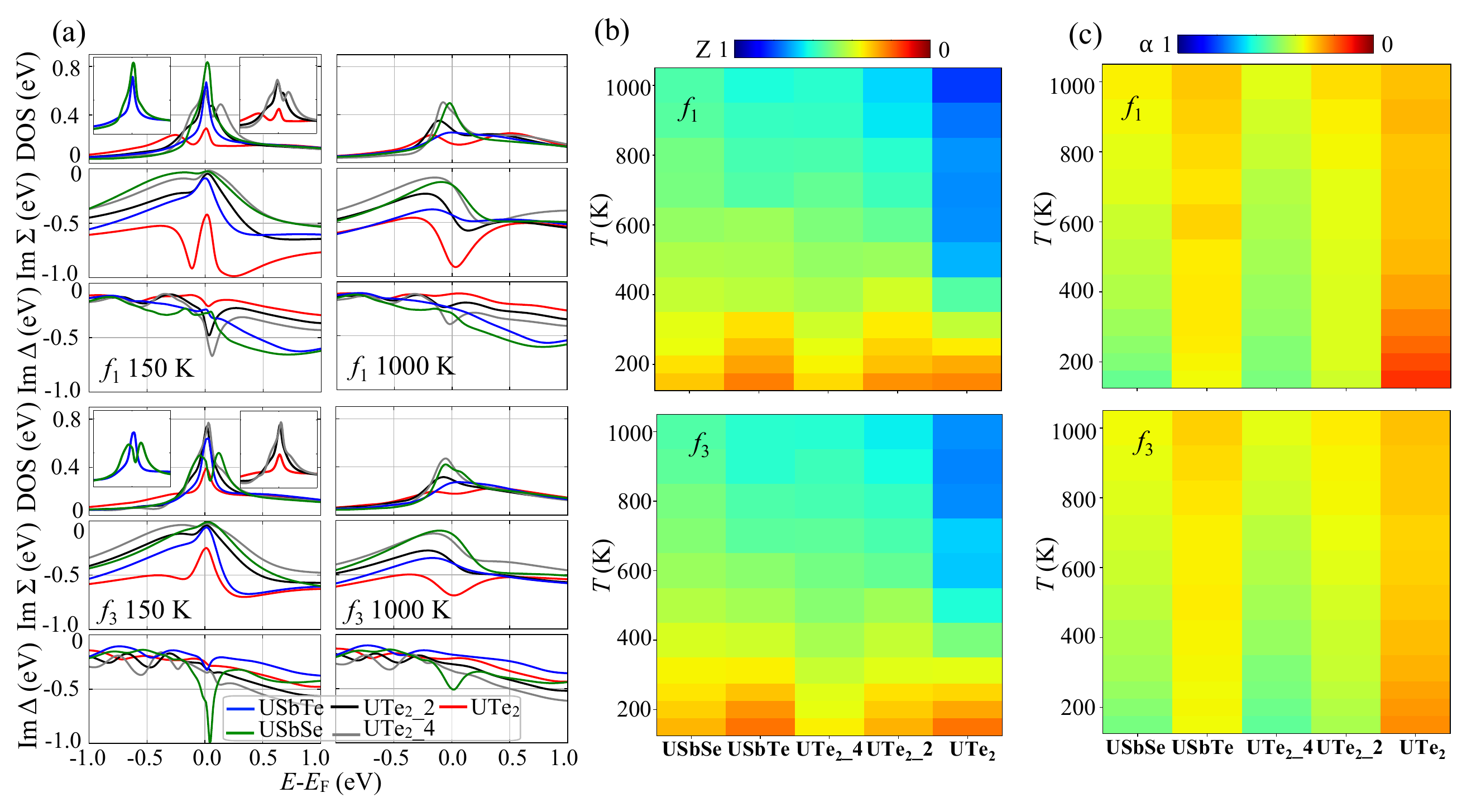}
\caption{\label{Fig_orbital} 
(a) DOS, imiginary part of self-energies and hybridization functions for $f_1$ and $f_3$.
Temperature dependent of (b) the quasi-particle weight $\it Z$ factors and (c) calculated $\alpha$ from the fitting of Im$\mathit{\Sigma}(\mathrm{i}\omega_{n}) \simeq -\mathit{\Gamma}+A(\mathrm{i}\omega_{n})^{\alpha}$, for $f_1$ and $f_3$.
In (a), inset shows magnified views of DOS in the vicinity of the Fermi level.
}
\end{figure*}
We sought to identify the origin of the duality by calculating orbital-projected values.
The six partially occupied U-5$f$ orbitals can be divided into three distinct groups, $f_{\alpha}$=$\{$$f_{1}$,$f_{6}$$\}$, $f_{\beta}$=$\{$$f_{2}$,$f_{5}$$\}$ and $f_{\gamma}$=$\{$$f_{3}$,$f_{4}$$\}$, based on their self-energy and hybridization functions, as illustrated in Fig.S3. 
Fig.~\ref{Fig_orbital} (a) displays the DOS, self-energies, and hybridization functions for $f_{1}$ and $f_{3}$, which correspond to $f_{\alpha}$ and $f_{\gamma}$, respectively.
Fig.~\ref{Fig_orbital} (b) shows the quasi-particle weight, $Z= 1/(1-\frac{\partial \Sigma(i\omega)}{\partial i\omega})$.  
Fig.~\ref{Fig_orbital} (c) shows $\alpha$ from the fitting of Im$\mathit{\Sigma}(\mathrm{i}\omega_{n}) \simeq -\mathit{\Gamma}+A(\mathrm{i}\omega_{n})^{\alpha}$. 
Deviation from the linear variation, i.e., non-Fermi liquid behavior, can be attributed to the freezing of localized spin moments ~\cite{philip_prl}.
The DOS, self-energy, hybridization function, Z and $\alpha$ for partially occupied six $f$ orbitals of all compounds were presented in Fig.S3, S4 and S5.

Figure~\ref{Fig_orbital} illustrates that UTe$_{2}$ has a higher $Z$ factor than other compounds at elevated temperatures, implying a feeble electron correlation and consequently a small quasi particle peak near the Fermi level.
As the temperature is decreased, the $Z$ factor decreases due to the emergence of incoherent $f$-$d$ Kondo hybridization and a large $f$ DOS in the vicinity of the Fermi level~\cite{byung_ute2}, as shown in Fig.~\ref{Fig_orbital} (a) and (b). UTe$_{2}$ also has a smaller $\alpha$ value compared to other materials, which is attributed to its more pronounced frozen magnetic moments.

Figure~\ref{Fig_orbital} shows that UTe$_{2}\_$4 and USbSe, which are in the coherent regime, have larger $\alpha$ values than UTe$_2$, indicating smaller magnetic moments.
At temperatures below 300 K, the $Z$ factors of UTe$_{2}\_$4 and USbSe are larger than that of UTe$_2$, indicating a suppression of electron correlations and an increase in the coherence energy of the electrons. This suggests a transition from an incoherent to a coherent behavior, from Curie-like to Pauli-like, at low temperatures~\cite{marianetti_prl2008}.
The transition is characterized by coherent $f$-$d$ hybridization, resulting in an increased quasi-particle peak near the Fermi level.
In contrast to UTe$_2$, UTe$_{2}\_$4 and USbSe exhibit orbital selective hybridization. The hybridization function of $f_{3}$ at the Fermi level displays strong divergence in contrast to $f_{1}$. This results in a two peak structure of the DOS in the vicinity of the Fermi level, unlike the DOS of $f_{1}$, as shown in Fig.~\ref{Fig_orbital} (a) and Fig.S6.
This suggests that the degree of itinerancy is dependent on the orbital configuration.
However, as shown in Fig.~\ref{Fig_orbital} (c) and Fig.S5, the orbital selective $f$-$d$ hybridization does not have a significant effect on $\alpha$ for the six partially occupied $f$ orbitals of USbSe. 
This indicates that the six $f$ orbitals contribute similarly to the magnetic moment.

Figure~\ref{Fig_dos} (e) shows the DOS of USbSe, which exhibits a kink in the vicinity of the Fermi level, indicating the presence of a VHS. This is due to the flat bands in the six partially occupied $f$ projected spectral functions, as seen in Fig.~\ref{Fig_dos} (f) and Fig.S7. 
The flat $f$ bands are subject to coherent $f$-$d$ hybridization, which is momentum dependent and forms partially flat bands along X-M-$\Gamma$ and R-A-Z symmetry lines, resulting in the emergence of VHS.
The momentum dependent $f$-$d$ hybridization in uranium-based Kondo lattices gives rise to a dual character of Pauli-like magnetic susceptibility and local magnetic moments of U-5$f$ electrons in the coherent regime. 
This duality is reminiscent of the behavior of $d$ electrons in Ni metallic systems, which act both as localized moments and itinerant contributions due to the VHS~\cite{hausoel_ncomm2017}.
Our results show that all partially filled U-5$f$ orbitals contribute to local magnetic moments, which can be distinguished from OSMP to explain the dual nature.
Table~\ref{table_occ} and Fig.~\ref{Fig_dos} demonstrate that the occupancy of U-5$f$ orbitals in the $J=5/2$ multiplet is away from half filling, indicating that the U-5$f$ in the presented compounds is not described by Mott-physics. This is further evidenced by the metallic U-5$f$ DOS shown in Fig.~\ref{Fig_dos}.

\textit{Investigating the magnetism of UTe$_2$ under pressure.}
To gain a better understanding of the origin of long-range magnetic ordering of UTe$_{2}$ at 1.5 GPa, we performed electronic structure calculations of UTe$_2$ in its orthorhombic phase with reduced volumes.
It was demonstrated in the preceding section that UTe$_2$ with reduced volumes are in a coherent state.
The temperature dependent $\chi_{J_Z}$ shown in Fig.~\ref{Fig_sus} (a) further supports the transition. The overall decrease in $\chi_{J_Z}$ with decreasing volume is in agreement with the experimental observation of a decrease in $\chi_{\mathrm{a}}$ under pressure~\cite{dexin_jpsj2021}.
Similar to USbSe, this coherent $f$-$d$ hybridization transition does not affect all $f$ bands in the reciprocal space, resulting in partially flat $f$ bands depending on momentum. This results in the emergence of VHS from the flat U-5$f$ bands along the L-T-W and R-X$_1$-Z symmetry lines, as shown in Fig.~\ref{Fig_dos} (c).
Our findings suggest that UTe$_2$ is in a coherent state under pressure, with a VHS causing long-range magnetic ordering at low temperatures.

We discuss other potential causes of long-range magnetic ordering in two distinct scenarios of UTe$_{2}$: fully-screened and under-screened.
In the fully-screened case, the net local magnetic moment is zero. To induce long-range magnetic ordering, the local magnetic moment must appear. Fig.~\ref{Fig_sus} (c) shows that the magnetic moment of U-5$f$ decreases with decreasing volume, which is in agreement with the experimental observation of a decrease in effective moments under pressure reported in Ref.~\cite{dexin_jpsj2021}.
Consequently, the reduced magnetic moment of U-5$f$ will remain fully-screened under pressure, with no impurities' magnetic moment to be coupled.

In the case of under-screened UTe$_{2}$, non-zero U-5$f$ magnetic moments can interact through either exchange or RKKY interaction.
It is reasonable to assume negligible direct exchange coupling between neighboring U-5$f$ moments, as the $f$-$f$ hybridization is negligible due to the small overlap of wave functions between U-5$f$. This is supported by our findings that $f$-$d$ hybridization is dominant at low energy.
We also suggest that RKKY interaction can be suppressed under pressure.
Doniach's work~\cite{doniach_phsica1977} suggests that the transition from an antiferromagnetic to a Kondo-like state is driven by the competition between the binding energy of a Kondo singlet, $W_{K}\sim N(0)^{-1}e^{-1/N(0)J_0}$, and that of the RKKY antiferromagnetic state, $W_{AF}\sim J_0^{2}N(0)$, where $J_0$ is exchange coupling constant between the localized moments and the conduction electrons and $N(0)$ is the density of conduction electron states.
When $J_0$ is below a certain threshold, the RKKY state is the most prominent, while above this, the Kondo singlet binding is the most influential.
The volume of UTe$_2$ was reduced, resulting in an increase in $J_{\textrm{H}}$ of U-6$d$ (see Fig.~\ref{Fig_u} (c)) due to the decrease in inter atomic distance.
These results suggest that the RKKY interaction is weakened by increased $J_0$ under pressure on UTe$_{2}$. Consequently, the RKKY interaction is not a viable explanation for the emergence of long-range magnetic ordering in UTe$_2$ under pressure.

\textit{Conclusion.} 
By altering the chemical composition or reducing the lattice parameters, uranium-based Kondo lattices are brought into a state of coherent $f$-$d$ hybridization, thus giving rise to the emergence of Bloch-like quasiparticles from the renormalized U-5$f$, which in turn bring about a duality of Pauli-like susceptibility and local magnetic moments through the van Hove singularity. This novel perspective on the magnetic properties of the Kondo lattice provides a new theoretical basis to explain the mechanism of long-range magnetic ordering in Kondo lattices.


\section*{Acknowledgments} 

We acknowledge the High Performance Computing Center (HPCC) at Texas Tech University for providing computational resources that have contributed to the research results reported within this paper.
M.H.K was supported by faculty startup funds from Texas Tech University.

\bibliography{ref}

\end{document}